\documentclass[aps,pra,reprint,showpacs,superscriptaddress,10pt]{revtex4-1}
\usepackage{amsmath,amssymb,amsfonts,graphicx,bm,longtable,dcolumn,color,mathrsfs,dsfont,lmodern,textcomp}
\usepackage[utf8]{inputenc}
\usepackage[T1]{fontenc}

\begin{document}

\title{Many-body radiative heat pumping}

\author{Riccardo Messina}\email{riccardo.messina@institutoptique.fr}
\affiliation{Laboratoire Charles Fabry, UMR 8501, Institut d'Optique, CNRS, Universit\'{e} Paris-Saclay, 2 Avenue Augustin Fresnel, 91127 Palaiseau Cedex, France}

\author{Philippe Ben-Abdallah}\email{pba@institutoptique.fr}
\affiliation{Laboratoire Charles Fabry, UMR 8501, Institut d'Optique, CNRS, Universit\'{e} Paris-Saclay, 2 Avenue Augustin Fresnel, 91127 Palaiseau Cedex, France}

\begin{abstract}
We introduce a local radiative heat-pumping effect between two bodies in a many-body system, obtained by periodically modulating both the temperature and the position of an intermediate object using an external source of energy. We show that the magnitude and the sign of energy flow can be tuned by changing the oscillation amplitude and dephasing of the two parameters. This many-body effect paves the way for an efficient and active control of heat fluxes at the nanoscale.
\end{abstract}

\maketitle

\section{Introduction}

Understanding and controlling heat exchanges at nanoscale is of tremendous importance both in fundamental and applied physics. During the last decades radiative heat exchanges between two bodies have been intensively studied~\cite{PoldervH,LoomisPRB94,JoulainSurfSciRep05,VolokitinRMP07,Song,Cuevas,HargreavesPLA69,KittelPRL05,NarayanaswamyPRB08,HuApplPhysLett08,ShenNanoLetters09,RousseauNaturePhoton09,OttensPRL11,KralikPRL12,vanZwolPRL12,SongNatureNano15,KimNature15,StGelaisNatureNano16,KloppstecharXiv,WatjenAPL16} and innovative solutions have been proposed~\cite{Thomas,Papadakis,Chen,Zhu,Ekeroth,Ilic,Messina_prb2017,Moncada,Cui,Ge,Ghanekar,Kou} for an active control of these transfers.

In recent theoretical works the possibility to pump heat in systems has been demonstrated~\cite{Segal,Shuttling,Segal2}. This pumping allows to transfer heat from the cold to the hot body by modulating, by means of external work, an intensive parameter such as the temperature or the chemical potential within the system. This effect is intimately related to a spatial symmetry breaking which allows to change the natural propagation direction of energy in the system using an external source of energy.

However, those pumping mechanisms require either the presence of a negative differential resistance~\cite{Shuttling,NDTR} or an exceptional point~\cite{Kottos,Berry} (i.e. a point at which two eigenstates coalesce under the variation of system parameters). The former case can be observed in systems made of phase-change materials~\cite{Mott,Baker,vanZwol} within temperature ranges around their critical temperature. As for the latter, it exists in non-Hermitian systems~\cite{Heiss} which exhibit a singularity of eigeinstates in the space of system parameters.

In the present letter we demonstrate the possibility to locally pump heat in many-body systems~\cite{PBA-PRL2011,KrugerPRB12,MessinaNdipoles,Messina-PRA2014,Latella,Pramod3body} by relaxing these constraints. More specifically, we show that in a Hermitian three-body system the modulation of both the position and the temperature of a body can induce a pumping effect between the two others ones. We show that the direction and amplitude of the energy flow can be significantly modulated by acting on the oscillation amplitude of temperature and position, as well as on their dephasing.

\section{Physical system and formalism}

\begin{figure}
	\includegraphics[width=0.47\textwidth]{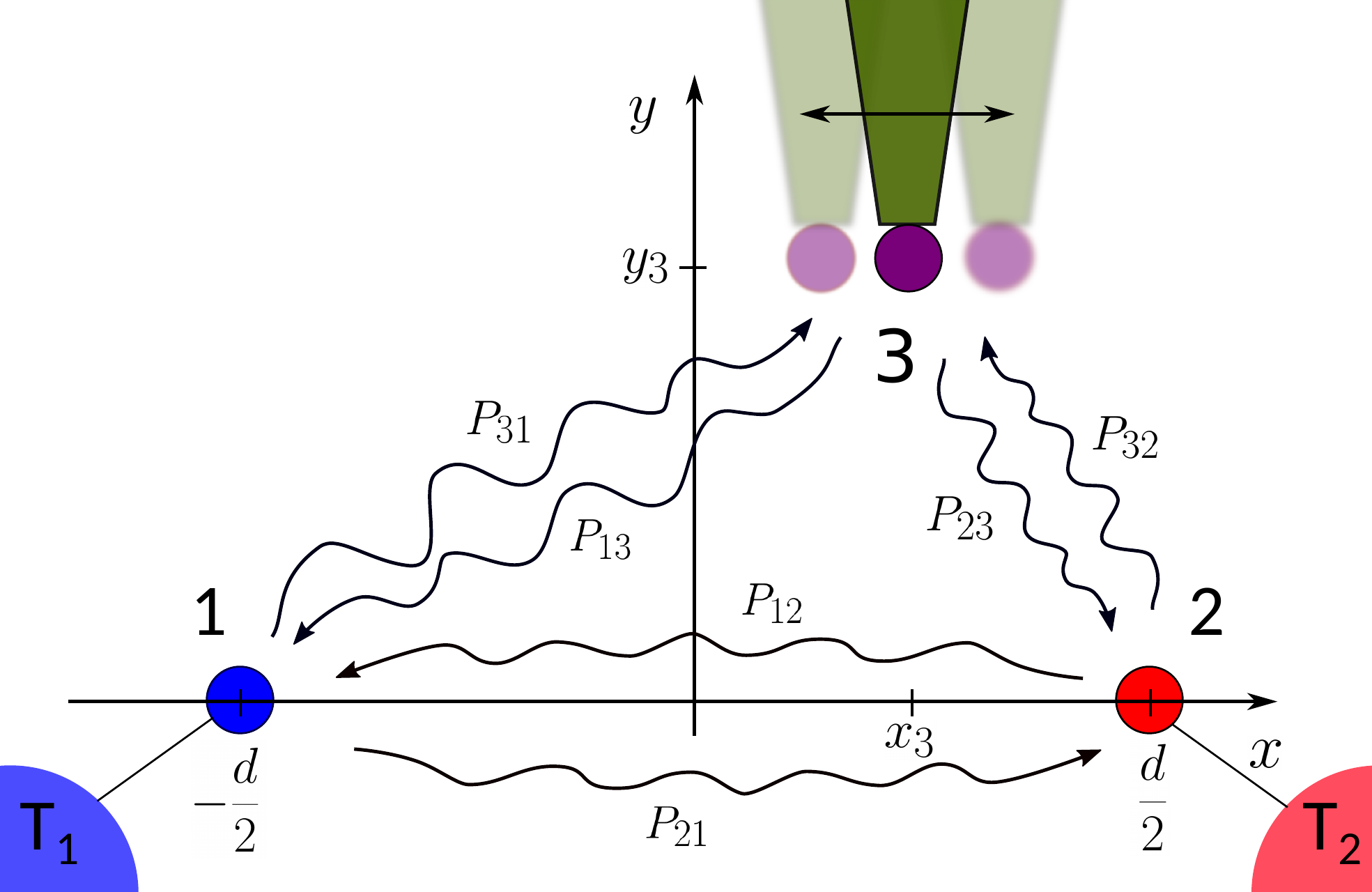}
	\caption{Geometry of the system. Particles 1 and 2 are placed at distance $d$ along the $x$ axis, while particle 3 has coordinates $x_3$ and $y_3$. The two individual contributions to the exchanged powers are shown. We also show a schematics of the dynamic oscillation of the $x$ coordinate of particle 3 procuced by a tip.}
	\label{fig:geometry}
\end{figure}

To introduce this effect let us consider consider the systesm sketched in Fig.~\ref{fig:geometry}. It is made of three spherical nanoparticles of radius $R$, labelled with indices 1, 2 and 3. Particles 1 and 2 constitute our reference system, on which the presence of particle 3 acts as an external source of modulation of the radiative power exchanged between 1 and 2. Particles 1 and 2 are assumed to be perfectly coupled with two thermostats at temperature $T_1$ and $T_2$, respectively. They have position $(-\frac{d}{2},0,0)$ and $(\frac{d}{2},0,0)$, where $d$ is the distance between the two particles, while particle 3 has position $(x_3,y_3,0)$ and temperature $T_3$.

Let us first introduce the formalism we are going to use to describe the radiative heat exchange within our three-particle system. We are going to calculate the energy transfer in the context of the dipolar approximation, according to which each particle is described purely in terms of a fluctuating and induced electric dipole. This approximation has proved to be valid when the center-to-center distance between the particles is larger than $4R$, $R$ being the radius of the particles~\cite{Philchain}. Under this approximation the optical response of each particle is described in terms of the electrical polarizability, which in the quasi-static approximation has the Clausius-Mossotti-like form
\begin{equation}
 \alpha(\omega)= 4 \pi a^3\frac{\epsilon(\omega)-1}{\epsilon(\omega)+2},
\end{equation}
where $\epsilon(\omega)$ is the dielectric permittivity of the material of the particle. 

The radiative power absorbed by each particle will be calculated by means of a fluctuational-electrodynamics approach, described in \cite{PBA-PRL2011,MessinaNdipoles}. In this theoretical framework the power absorbed by particle $i$ (Latin indices refer to particles and take values 1, 2 and 3) reads 
\begin{equation}\label{eq:Pi}P_i=\int_0^{+\infty}\frac{d\omega}{2\pi}\hbar\omega\sum_{j\neq i}\frac{4\chi_i\chi_j}{|\alpha_i|^2}n_{ji}(\omega)\text{Tr}\Bigl(\mathds{T}^{-1}_{ij}\mathds{T}^{-1\dag}_{ji}\Bigr),\end{equation}
where $n_{ij}(\omega)=n(\omega,T_i)-n(\omega,T_j)$ is the difference of Bose--Einstein distributions 
\begin{equation}
 n(\omega,T) = \biggl[\exp\biggl(\frac{\hbar\omega}{k_B T}\biggr) - 1\biggr]^{-1},
\end{equation}
calculated at the temperature of particles $i$ and $j$. Moreover, in Eq.~\eqref{eq:Pi} we have introduced the susceptibility $\chi_i$ of each particle, defined as~\cite{abajo}
\begin{equation}
 \chi_i=\text{Im}(\alpha_i)-\frac{\omega^3}{6\pi c^3} |\alpha_i|^2,
\end{equation}
and $\mathbb{T}$ is a $3N\times3N$ block matrix defined in terms of the $(i,j)$ $N\times N$ sub-matrices ($i,j=1,\dots,N$)
\begin{equation}
 \mathds{T}_{ij} = \delta_{ij}\mathds{1}-(1-\delta_{ij})\frac{\omega^2}{c^2}\alpha_i\mathds{G}_{ij},
\end{equation}
$\mathds{G}_{ij}$ being the Green function in vacuum evaluated at the coordinates of dipoles $i$ and $j$.

\section{Results}

\subsection{Pumping effect}

In analogy with the electrical pumping that transfers charge against a bias voltage by modulating system parameters~\cite{Thouless}, we demonstrate here the possibility of tailoring the net power absorbed by the two particles 1 and 2 by means of periodic variations of physical properties and interaction strengths with a third body. More specifically, we are going to consider both the variation of the temperature $T_3$ of this body around its equilibrium value $T_{3,\rm{eq}}$ and the oscillation of its coordinate $x_3$ along the axis between the two thermostated particles. In a scanning thermal microscopy setup this corresponds to a transerse tapping mode. To demonstrate the potential of such modulations we start by describing how the power absorbed by particle 1 depends on this change of parameters. To this aim we develop this power using a Taylor expansion, at the second order, with respect to these parameters around $x_3=0$ and $T_3=T_{3,\text{eq}}$. This formally reads
\begin{equation}\begin{split}
 P_1 &\simeq P_1(0,T_{3,\text{eq}}) + \frac{\partial P_1}{\partial x_3}x_3 + \frac{\partial P_1}{\partial T_3}(T_3 - T_{3,\text{eq}})\\
 &\,+ \frac{1}{2} \frac{\partial^2 P_1}{\partial x_3^2}x_3^2 + \frac{1}{2} \frac{\partial^2 P_1}{\partial T_3^2}(T_3 - T_{3,\text{eq}})^2\\
 &\,+ \frac{\partial^2 P_1}{\partial x_3\partial T_3}x_3(T_3 - T_{3,\text{eq}}).
\end{split}\end{equation}
In the specific case where particles 1 and 2 are held at the same temperature $T_1=T_2=T_{eq},$, then $T_{3,\rm{eq}}=T_{eq}$ and the net power absorbed by particles 1 and 2 vanishes (i.e. $P_i=0$, $i=1,2$) for any value of $x_3$. As a consequence, the first, second and fourth terms are identically equal to zero. Moreover, with the periodic modulation 
\begin{equation}\label{eq:Tx3t}
T_3(t)=T_{3,\rm{eq}} + \Delta T\sin(\omega t),\,\,x_3(t)=\Delta x\sin(\omega t+\phi),
\end{equation}
of parameters the third term has a vanishing average over the oscillation period so that the average power reads
\begin{equation}\label{eq:P1ave}
\langle P_1\rangle \simeq \frac{\Delta T}{2}\Bigl(\Delta x\frac{\partial^2P_1}{\partial x_3\partial T_3}\cos\phi + \frac{\Delta T}{2}\frac{\partial^2 P_1}{\partial T_3^2}\Bigr),
\end{equation}
 $\langle .\rangle$ denoting the time averging. The second term of this equation exists also when the temperature $T_3$ is the only varying parameter. In this case, as discussed in \cite{Shuttling}, in the absence of a negative thermal differential resistance, this contribution is necessarily positive. On the contrary, it is manifest that the magnitude and sign of the first term can be easily modulated simply by acting on the dephasing $\phi$ between $x_3$ and $T_3$. This gives a clear evidence of the remarkable advantage of exploiting the simultaneous variation of two parameters instead of one as it is done in the classical shuttling mechanism~\cite{Shuttling}. 

For concretness , we consider a system made with three particles of silicon carbide (SiC)~\cite{Palik}, for which the dielectric permittivity is well described by a Drude-Lorentz expression $\varepsilon(\omega)=\varepsilon_\infty (\omega^2_L-\omega^2-i\Gamma\omega)/(\omega^2_T-\omega^2-i\Gamma\omega)$, where $\varepsilon_\infty=6.7$, $\omega_L=1.83\times 10^{14}\,$rad/s, $\omega_T=1.49\times 10^{14}\,$rad/s, and $\Gamma=8.97\times 10^{11}\,$rad/s. This model predicts the existence of a phonon-polariton resonance for a spherical SiC nanoparticle at a frequency $\omega=1.755\times10^{14}\,$rad/s, giving $\lambda\simeq11\,\mu$m, close to the emission peak of a blackbody at 300\,K.
We also consider a period of oscillation of parameters of 1\,s ($\omega=2\pi/1\,$s), which allow us to assume an adiabiatic variation of both $T_3$. Moreover, we take $\Delta T=5\,$K, $\Delta x=2R=100\,$nm and $\phi=0$, with $d=12R=600\,$nm and $y_3=d/2=300\,$nm. The power absorbed by the three particles as a function of time during one period is shown in Fig.~\ref{fig:pumping}(a). The analysis of $P_1$ and $P_2$ proves how we are able to exploit at the same time the control of the sign of the flux through $T_3$ and the strong dependence of the flux on the distance through the coordinate $x_3$. More specifically, during the first half of the period the temperature $T_3$ is above the equilibrium temperature, resulting in a power absorption on both particles 1 and 2, much higher on particle 2 because of the positive coordinate $x_3$. The opposite scenario happens during the second half of the period. As a result, the average powers transferred to particles 1 and 2 during one period equal $(\langle P_1\rangle,\langle P_2\rangle)=(-2.18,2.30)\times10^{-14}\,$W. We are thus in the presence of a heat-pump effect, where the energy transfer between particles 1 and 2, at thermal equilibrium between each other, is produced by the presence and modulation of properties of a third particle introduced in the system. We observe that, for the parameters chosen here, the exchanged power is the one the two particles would exchange in a two-body configuration for a temperature difference $\Delta T=T_2-T_1\simeq20\,$K, while the value of $T_3$ oscillates over $\Delta T/2=10$\,K.

\begin{widetext}
	\begin{center}
		\begin{figure}[t!]
			\includegraphics[width=0.3\textwidth]{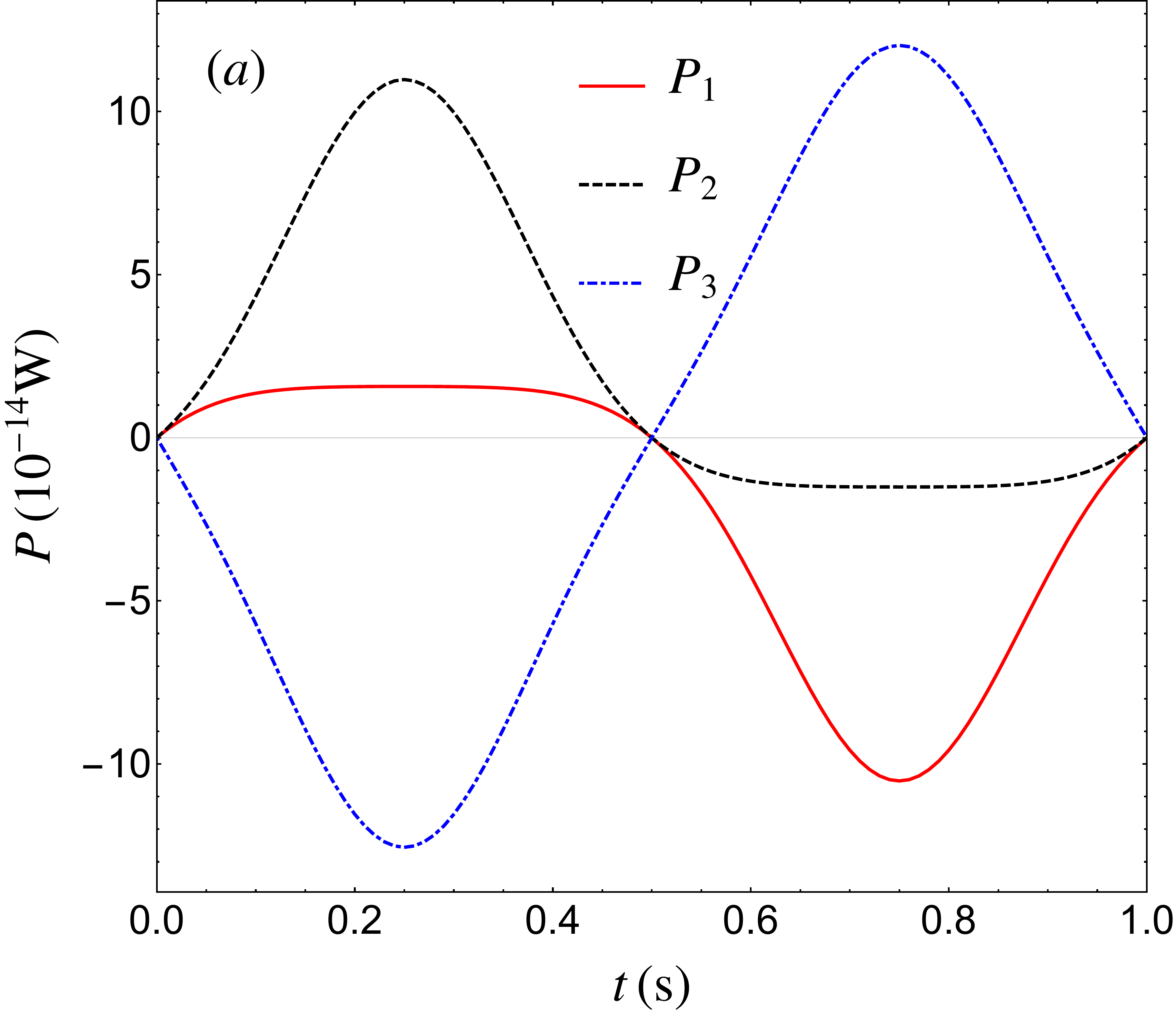}\hspace{.5cm}\includegraphics[width=0.32\textwidth]{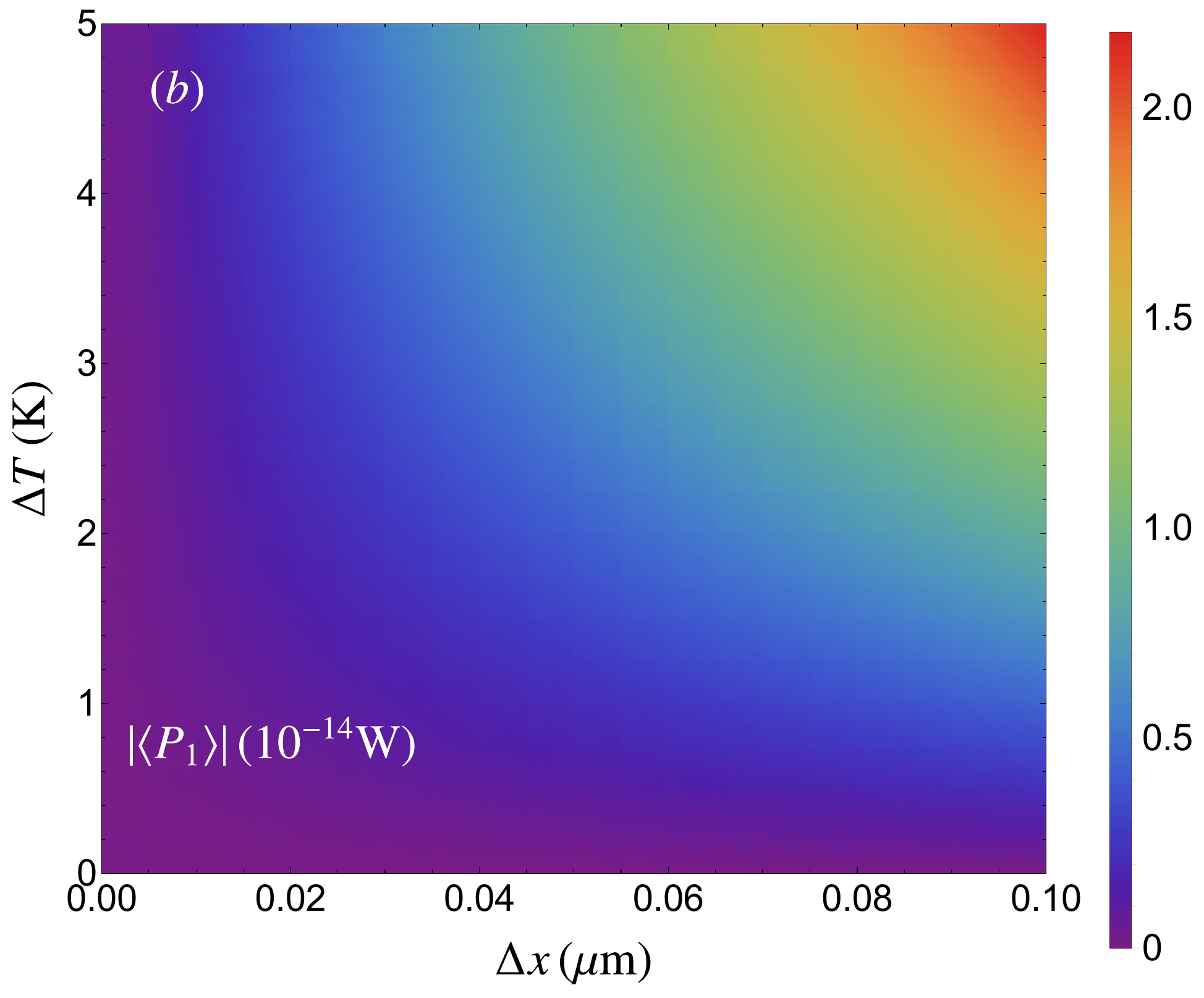}\hspace{.5cm}\includegraphics[width=0.3\textwidth]{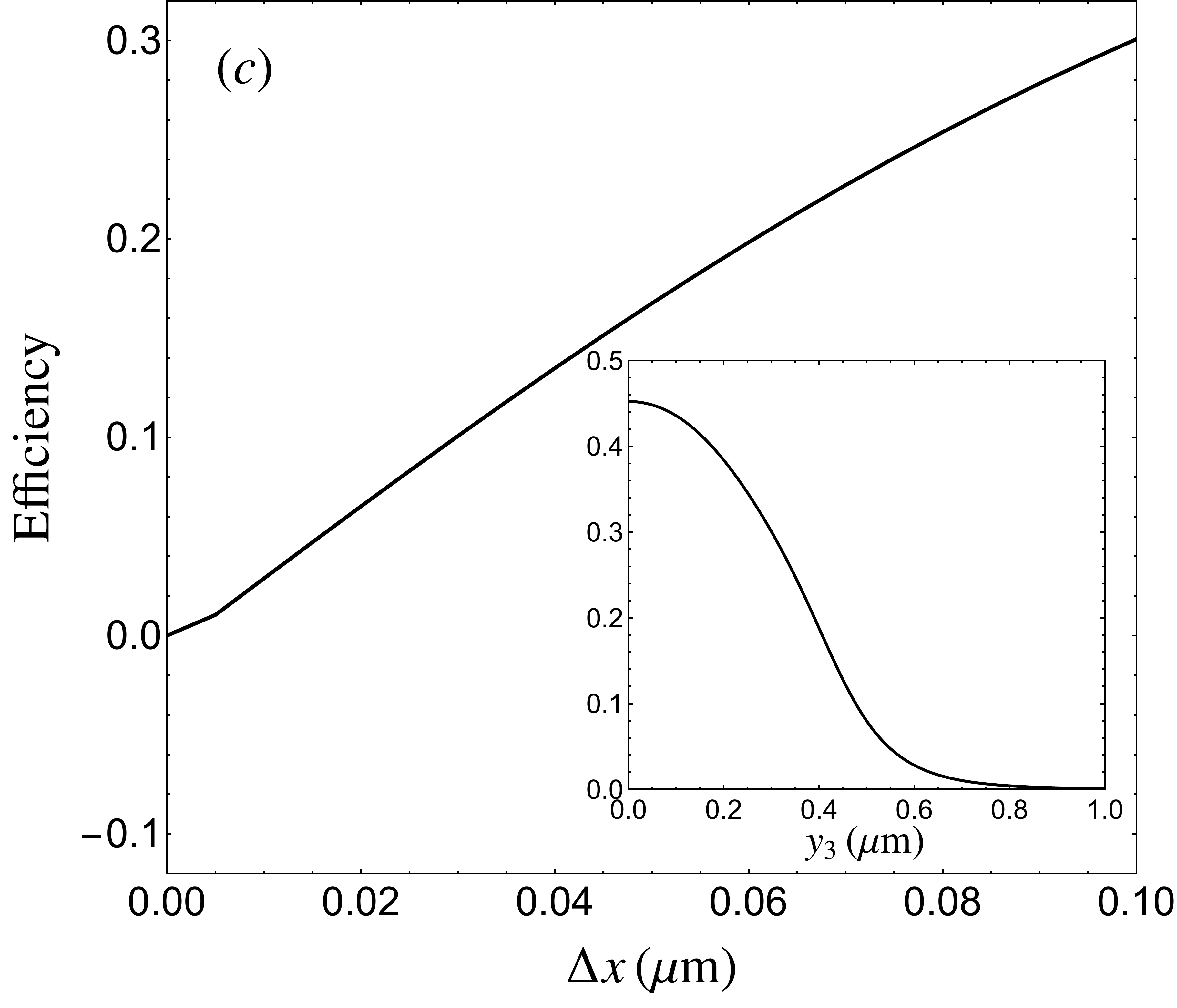}
			\caption{(a) Powers absorbed by particles 1 (solid red line), 2 (dashed black line) and 3 (blue dot-dashed line) as a function of time for the periodic variation given by Eq.~\eqref{eq:Tx3t} with $\omega = 2\pi/1\,$s, $\phi = 0$, $\Delta x=2R=100\,$nm and $\Delta T=5\,$K. (b) Absolute value of the average power absorbed by particle 1 over one period as a function of the displacement and temperature oscillation amplitudes $\Delta x$ and $\Delta T$, respectively. (c) In the main part of the figure, efficiency of the pumping effect as a function of the diplacement oscillation amplitude $\Delta x$ for $y_3=300\,$nm. In the inset, efficiency as a function of $y_3$ for $\Delta x=2R=100\,$nm.}
			\label{fig:pumping}
		\end{figure}
	\end{center}
\end{widetext}

It is interesting at this stage to understand how the variation of the temperature and displacement oscillation amplitudes, $\Delta T$ and $\Delta x$ respectively, affect the net power received by particles 1 and 2. This is shown in Fig.~\ref{fig:pumping}(b), where the absolute value of $E_1$ is represented as a function of $\Delta T$ and $\Delta x$. We clearly see that not only does this energy increase with respect to both amplitudes, but also that it vanishes when $\Delta T$ or $\Delta x$ go to zero, confirming that the simultaneous variation of two parameters is essential to produce a pumping effect.

In order to better characterize this pumping effect, it is important to define its efficiency in terms of ratio between the average power extracted from particle 1 or absorbed by particle 2 and the external energy needed to produce the periodic variation of $x_3$ and $T_3$. As far as the temperature variation is concerned we start by considering the equation governing the time variation of the internal energy of particle 3
\begin{equation}
\rho_3 C_3 V_3 \frac{d T_3(t)}{dt}= P_3 + P_\text{ext,th},
\end{equation}
wher $\rho_3$, $C_3$, and $V_3$ represent its mass density, heat capacity and volume, respectively, while $P_3$ is the radiative power transfer to particle 3 due to the presence of particles 1 and 2, whereas $P_\text{ext}$ is the external power injected in the system. Since the temperature time dependence is fixed by Eq.~\eqref{eq:Tx3t} and the radiative power $P_3$ can be calculated by using the formalism introduced above, we can deduce the instantaneous power $P_\text{ext}$ and its average over one period. Concerning the mechanical oscillation, in order to give a realistic model of our system we assume that a spherical particle (particle 3) is attached to a cantilever and describe this system as a forced harmonic oscillator~\cite{Springer}. The average power absorption during an oscillation cycle reads
\begin{equation}
\langle P_\text{ext,mech}\rangle = \frac{1}{2}kX_0\Delta x\omega\sin\varphi,
\end{equation}
where $k=3EI/L^3$ is the stifness of the cantilever ($E$ being its Young modulus, $I$ its second moment of area and $L$ its length), $X_0$ the oscillation amplitude of the oscillating applied force, $\omega$ its frequency, and $\varphi$ the dephasing between it and the displacement. By solving the mechanical problem, $X_0$ can be connected to $\Delta x$, and assuming for simplicity that the cantilever is made of SiC as well, we conclude that the absorbed power is approximately given by~\cite{Springer}
\begin{equation}
\langle P_\text{ext,mech}\rangle \simeq \sqrt{\rho_3 E} \Delta x^2 \frac{\omega^2 d^3}{Q L},
	\end{equation}
where $d$ is the side of the square section of the cantilever and $Q$ its quality factor ($\simeq 100$ in air). For $d=1\,$mm and $L=10\,$cm, this power is of the order of $10^{-15}\,$W, and it can be made even smaller by acting on the dimensions and material of the cantilever. For this reason, we will neglect from now on this contribution to the external power and define the efficiency as $\eta = |\langle P_1\rangle|/P_\text{ext,th}$. We have observed that the calculated efficiency depends very weakly on the temperature amplitude $\Delta T$. This is not surprising, since for small temperature differences all the energy exchanges are linear with respect to $\Delta T$ and so is their ratio as well. On the contrary, the dependence of the efficiency on $\Delta x$ is shown in Fig.~\ref{fig:pumping}(c). It shows that the efficiency (as the exchanged powers) goes to zero for $\Delta x$ approaching zero, and it reaches a remarkable value around 30\% for $\Delta x=2R=100\,$nm. In order to further confirm the near-field nature of this pumping effect we show in the inset of Fig.~\ref{fig:pumping}(c) the effiency as a function of $y_3$ for $\Delta x=100\,$nm. We see that the efficiency is maximized, as expeced, in the case $y_3=0$ (corrisponding to a linear chain), where it reaches 46\%, while decreases quickly and tends to zero, even for non-vanishing $\Delta x$ and $\Delta T$, when increasing $y_3$ above 500\,nm.

\begin{figure}
	\includegraphics[width=0.47\textwidth]{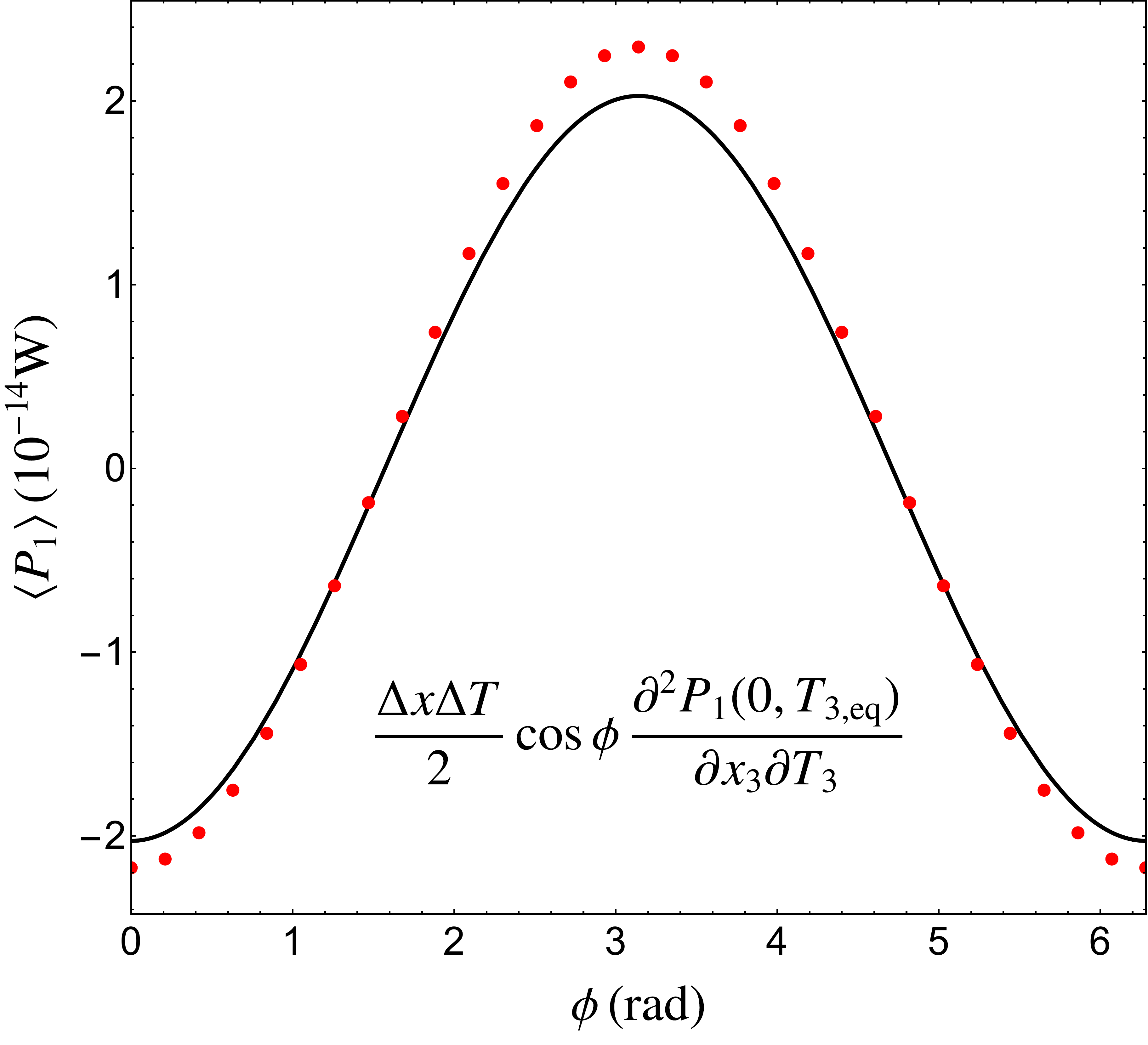}
	\caption{Average power absorbed by particle 1 as a function of the dephasing $\phi$ between the temperature and position variations defined in Eq.~\eqref{eq:Tx3t}. The red points are the numerical resuts, compared to the black solid line associated with the first term of Eq.~\eqref{eq:P1ave}.}
	\label{fig:P1phi}
\end{figure}

We conclude by showing the effect of the dephasing $\phi$ on the average absorbed power $P_1$. Referring to Eq.~\eqref{eq:P1ave}, we start by observing that the values of these two derivatives can be deduced numerically. In our specific case, we have $\partial^2P_1/\partial x_3\partial T_3=-8.11\times10^{-8}\,\text{W}\,\text{m}^{-1}\,\text{K}^{-1}$ and $\partial^2 P_1/\partial T_3^2=7.01\times 10^{-17}\,\text{W}\,\text{K}^{-2}$ so that the second term proves to be negligible with respect to the first one. Moreover, in the approximation of weak temperature difference we can linearize the power transferred to particle 1 as $P_1\simeq G_{12}(T_2-T_1)+G_{13}(T_3-T_1)$, where the first term vanishes since $T_1=T_2$. As a consequence, we have $\partial^2P_1/\partial x_3\partial T_3\simeq\partial G_{13}/\partial x_3$, which is negative since the power transferred from particle 3 to particle 1 clearly decreases when increasing $x_3$, thus the distance between the two particles. Figure \ref{fig:P1phi} shows the pumped power $P_1$ as a function of $\phi$, confirming at the same time the modulation in amplitude and sign due to the dephasing $\phi$ and the good agreement between the numerical results (red points) and the analytical approximation given by the first term in Eq.~\eqref{eq:P1ave}.

\subsection{Control of the direction of the flux}

In the above discussions, particles 1 and 2 were in relative equilibrium. For practical applications it is also important to investigate the situation where a temperature gradient exists between these two particles.

\begin{widetext}
\begin{center}
\begin{figure}
	\includegraphics[width=0.4\textwidth]{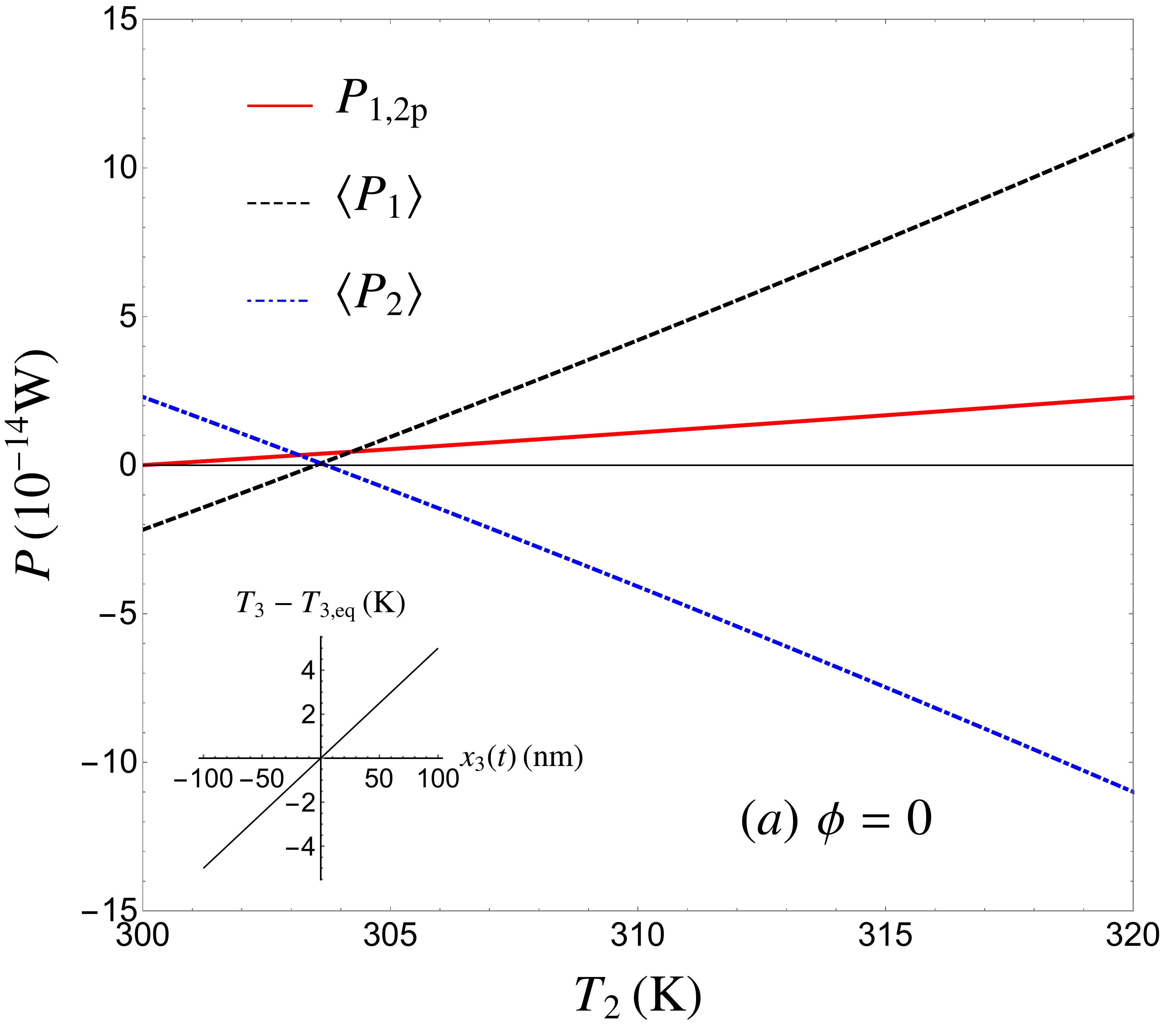}\hspace{1.5cm}\includegraphics[width=0.4\textwidth]{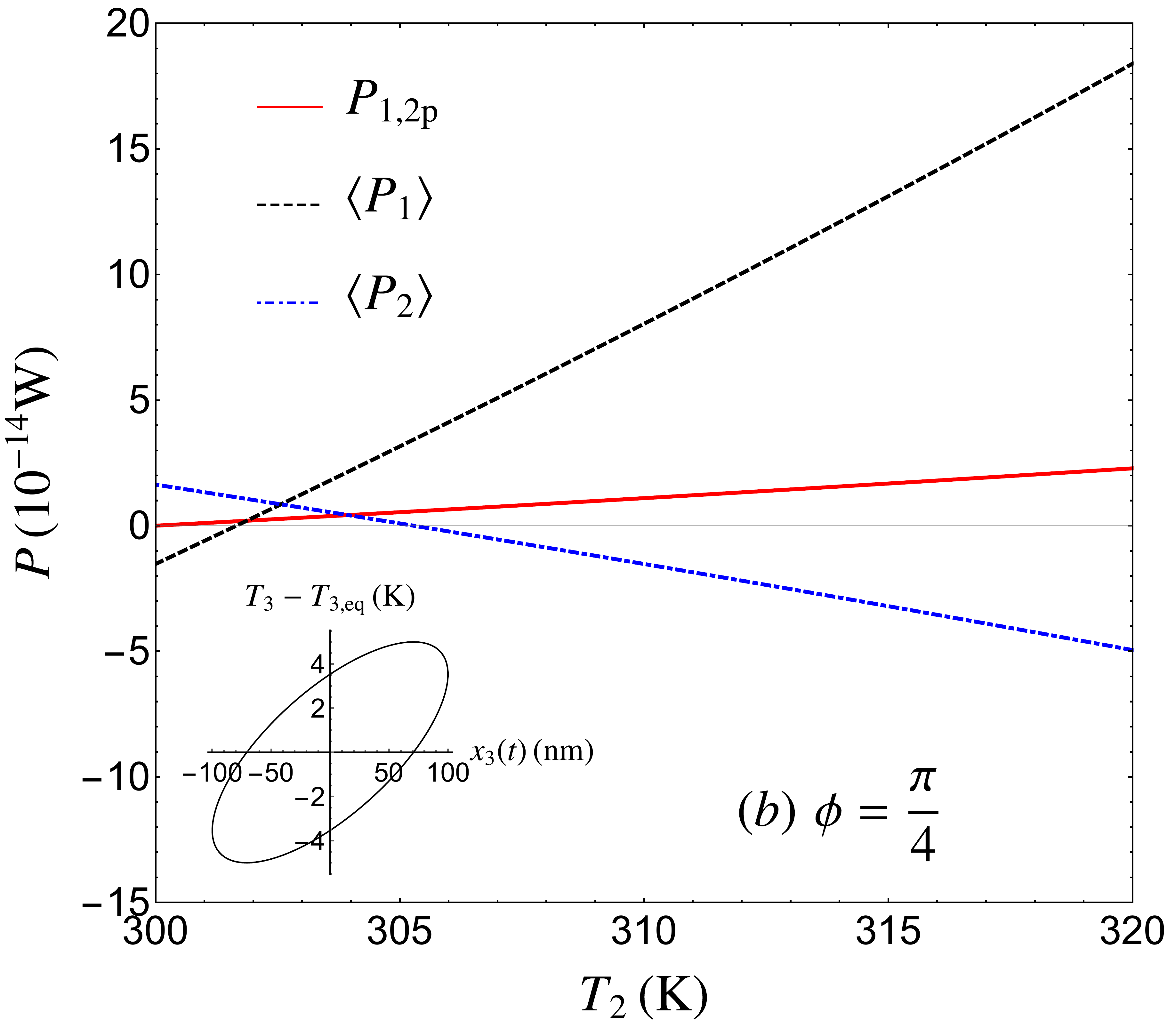}\\
	\includegraphics[width=0.4\textwidth]{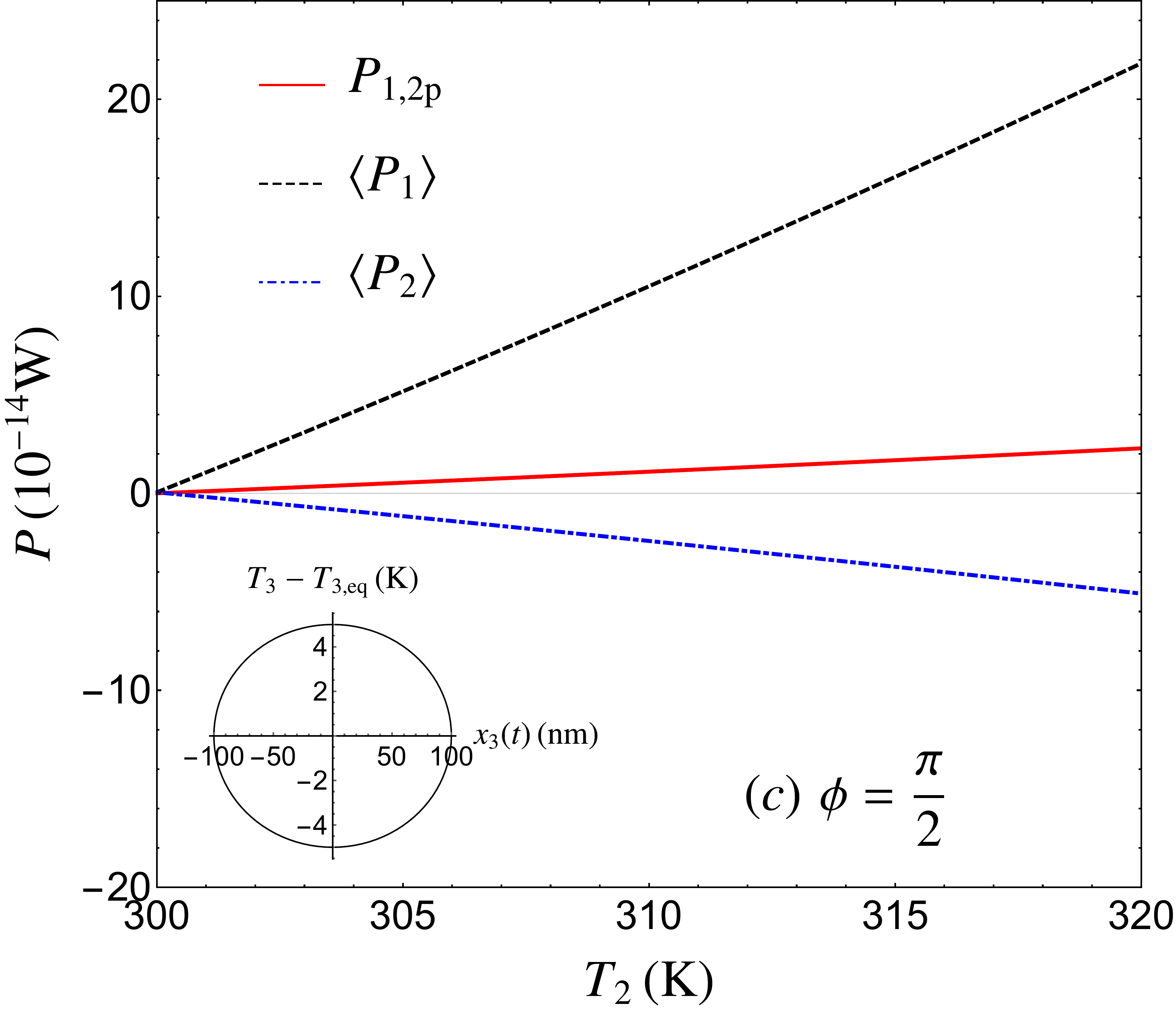}\hspace{1.5cm}\includegraphics[width=0.4\textwidth]{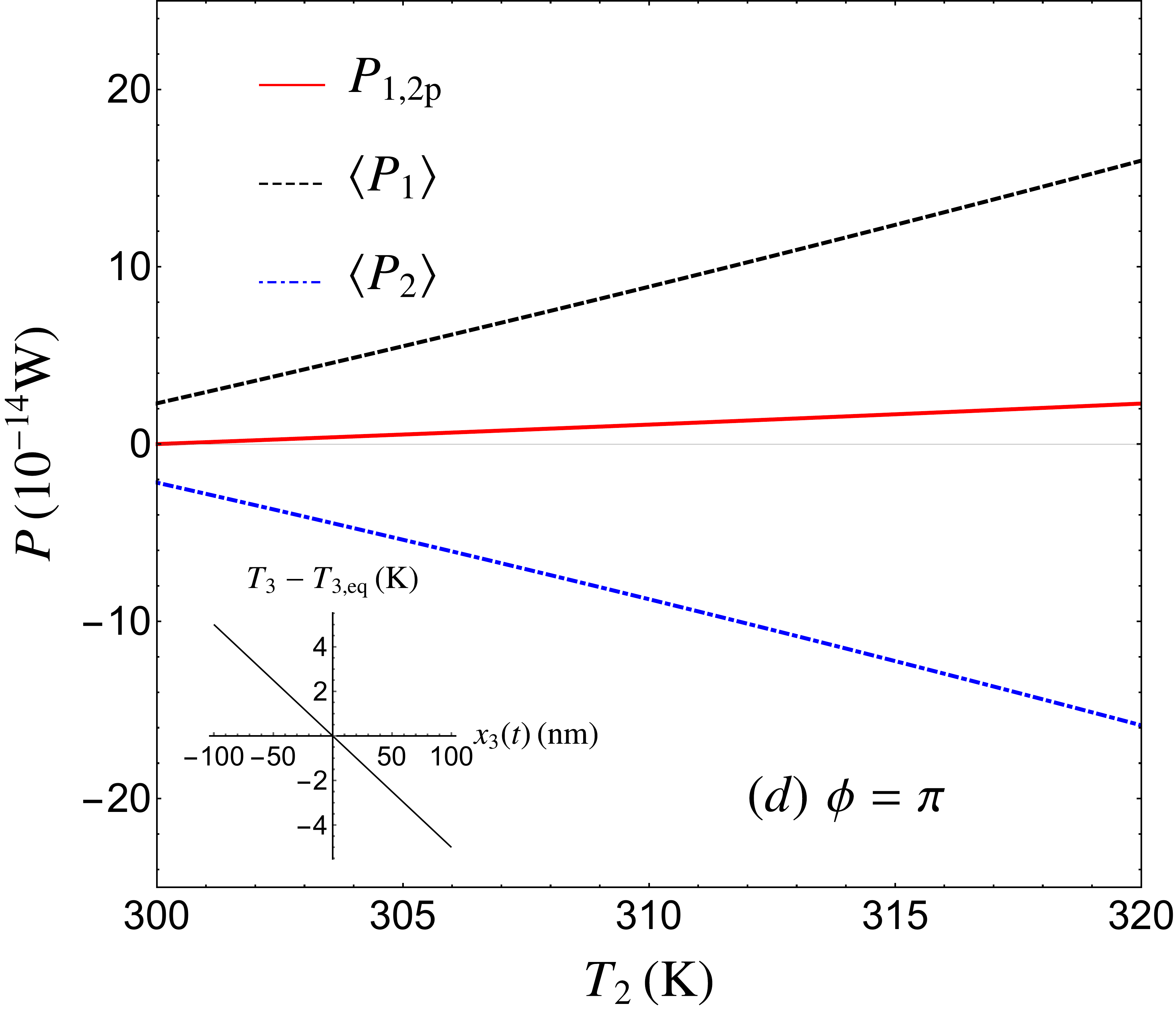}\\
	\caption{Average power absorbed by particle 1 (black dashed line) and particle 2 (blue dot-dashed line) compared to the two-body power absorbed by particle 1 (red solid line). We have $\Delta x =2R=100\,$nm, $\Delta T=5\,$K, $d=12R=600\,$nm and $y_3=d/2=300\,$nm. The four panels correspond to different dephasings $\phi$ between $x_3$ and $T_3$: (a) $\phi = 0$, (b) $\phi = \pi/4$, (c) $\phi = \pi/2$, (d) $\phi = \pi$.}
	\label{fig:fT2}
\end{figure}
\end{center}
\end{widetext}

This situation is depicted in Figs.~\ref{fig:fT2}, in the case where $T_2$, vary above $T_1$ and $T_1=300$K . In these plots the red solid lines show as a reference the power exchanged between particle 1 and 2 in the absence of particle 3. The black dashed line and the blue dot-dashed one show, respectively, the average power absorved by particles 1 and 2 over a periodic oscillation of $x_3$ and $T_3$. The parameters are $\Delta x =2R=100\,$nm, $\Delta T=5\,$K, $d=12R=600\,$nm and $y_3=d/2=300\,$nm. Moreover, in order to demonstrate the importance of the role played by the dephasing $\phi$ between $x_3$ and $T_3$, the four cases $\phi=0,\pi/4,\pi/2,\pi$ are shown. We stress that, for each value of $T_2$, $T_3$ oscillates around its equilibrium value defined by $P_3=0$, which depends on $T_2$ and is always close, for small temperature differences, to the average $(T_1+T_2)/2$.

When the temperature gradient vanishes and the dephasing $\phi=0$ we recover the pumping shown previously, with average absorbed powers $(\langle P_1\rangle,\langle P_2\rangle)=(-2.18,2.30)\times10^{14}\,$W. This effect is still present but lower for $\phi=\pi/4$, while no pumping effect happens for $\pi/2$. Finally, for $\phi=\pi$, the power absorbed is different from zero, but goes in the opposite direction with respect to $\phi=0$, coherently with the simple analytical description given in Eq.~\eqref{eq:P1ave}.

In presence of temperature gradient it is relevant to distinguish the cases $\phi=0,\pi/4$ from the cases $\phi=\pi/2,\pi$. As a matter of fact, for $\phi=0,\pi/4$, we observe a region of $T_2$ [up to values around 304\,K (302,\,K) for $\phi=0$ ($\phi=\pi/4$)] for which the exchanged power remarkably goes in the direction opposite to the temperature difference. This effect gets weaker when increasing $T_2$ and we also highlight the complementary result that, for $\phi=0$ and $T_2\simeq304\,$K, particles 1 and 2 are thermally isolated even though they are not at thermal equilibrium. On the contrary, for $\phi=\pi/2,\pi$ the exchanged power always goes in the direction of the temperature difference between particles 1 and 2, but the amplitude of this power can be clearly modulated by acting on the dephasing $\phi$. Moreover, for any dephasing considered here, the exchanged power is significantly amplified with respect to the two-body value. This analysis shows that our scheme and the introduction and active control of a third particle in the system can be used to tailor in several different ways the heat flux between two bodies.

\section{Conclusions}

In conclusion, we have demonstrated that a heat pumping current can be locally induced between two bodies within a many-body system by periodically varying independently the temperature and the location of an intermediate object. By changing the dephasing between these two parameters we have shown the possibility to control the magnitude and signed of pumped flux. Unlike previously highlighted effects, our pumping effect does not require the presence of channels with a negative differential thermal resistance within the system neither the existence of exotic topological characteristics such as exceptional points. While in the present work we have limited our study to the variation of temperature and position, the same effect could be realized by varying other intensive parameters, such as the chemical potential, or external fields, for example by exploiting magneto-optical or ferroelectric properties.

As far as applications are concerned, the efficiency of the observed effect makes it promising for an active control of heat currents at the nanoscale, in the adiabatic limit. It could also find applications in the field of energy conversion.


\begin{thebibliography}{99}
\bibitem{PoldervH}D. Polder and M. van Hove, \emph{Theory of Radiative Heat Transfer between Closely Spaced Bodies}, Phys. Rev. B \textbf{4}, 3303 (1971).
\bibitem{LoomisPRB94}L.~J. Loomis and H.~J. Maris, \emph{Theory of heat transfer by evanescent electromagnetic waves}, Phys. Rev. B \textbf{50}, 18517 (1994).
\bibitem{JoulainSurfSciRep05}K. Joulain, J.-P., Mulet, F. Marquier, R. Carminati, and J.-J. Greffet, \emph{Surface electromagnetic waves thermally excited: Radiative heat transfer, coherence properties and Casimir forces revisited in the near field}, Surf. Sci. Rep. \textbf{57}, 59 (2005).
\bibitem{VolokitinRMP07}A.~I. Volokitin and B. N. Persson, \emph{Near-field radiative heat transfer and noncontact friction}, Rev. Mod. Phys. \textbf{79}, 1291 (2007).
\bibitem{Song}B. Song, A. Fiorino, E. Meyhofer, and P. Reddy, \emph{Near-field radiative thermal transport: From theory to experiment}, AIP Advances \textbf{5}, 053503 (2015).
\bibitem{Cuevas}J.~C. Cuevas and F.~J. García-Vidal, \emph{Radiative Heat Transfer}, ACS Photonics \textbf{5}, 3896 (2018).
\bibitem{HargreavesPLA69}C. Hargreaves, \emph{Anomalous radiative transfer between closely-spaced bodies}, Phys. Lett. A \textbf{30}, 491 (1969).
\bibitem{KittelPRL05}A. Kittel, W. M\"{u}ller-Hirsch, J. Parisi, S.-A. Biehs, D. Reddig, and M. Holthaus, \emph{Near-Field Heat Transfer in a Scanning Thermal Microscope}, Phys. Rev. Lett. \textbf{95}, 224301 (2005).
\bibitem{NarayanaswamyPRB08}A. Narayanaswamy, S. Shen, and G. Chen, \emph{Near-field radiative heat transfer between a sphere and a substrate}, Phys. Rev. B \textbf{78}, 115303 (2008).
\bibitem{HuApplPhysLett08}L. Hu, A. Narayanaswamy, X. Chen, and G. Chen, \emph{Near-field thermal radiation between two closely spaced glass plates exceeding Planck’s blackbody radiation law}, Appl. Phys. Lett. \textbf{92}, 133106 (2008).
\bibitem{ShenNanoLetters09}S. Shen, A. Narayanaswamy, and G. Chen, \emph{Surface Phonon Polaritons Mediated Energy Transfer between Nanoscale Gaps
}, Nano Letters \textbf{9}, 2909 (2009).
\bibitem{RousseauNaturePhoton09}E. Rousseau, A. Siria, G. Joudran, S. Volz, F. Comin, J. Chevrier, and J.-J. Greffet, \emph{Letter
Published: 23 August 2009
}, Nature Photon. \textbf{3}, 514 (2009).
\bibitem{OttensPRL11}R.~S. Ottens, V. Quetschke, S. Wise, A.~A. Alemi, R. Lundock, G. Mueller, D.~H. Reitze, D.~B. Tanner, and B.~F. Whiting, \emph{Near-Field Radiative Heat Transfer between Macroscopic Planar Surfaces}, Phys. Rev. Lett. \textbf{107}, 014301 (2011).
\bibitem{KralikPRL12}T. Kralik, P. Hanzelka, M. Zobac, V. Musilova, T. Fort, and M. Horak, \emph{Strong Near-Field Enhancement of Radiative Heat Transfer between Metallic Surfaces}, Phys. Rev. Lett. \textbf{109}, 224302 (2012).
\bibitem{vanZwolPRL12}P.~J. van Zwol, L. Ranno, and J. Chevrier, \emph{Tuning Near Field Radiative Heat Flux through Surface Excitations with a Metal Insulator Transition}, Phys. Rev. Lett. \textbf{108}, 234301 (2012).
\bibitem{SongNatureNano15}B. Song, Y. Ganjeh, S. Sadat, D. Thompson, A. Fiorino, V. Fern\'{a}ndez-Hurtado, J. Feist, F.~J. Garcia-Vidal, J.~C. Cuevas, P. Reddy, and E. Meyhofer, \emph{Enhancement of near-field radiative heat transfer using polar dielectric thin films}, Nat. Nanotechnol. \textbf{10}, 253 (2015).
\bibitem{KimNature15}K. Kim, B. Song, V. Fern\'{a}ndez-Hurtado, W. Lee, W. Jeong, L. Cui, D. Thompson, J. Feist, M.~T. Homer Reid, F.~J. Garcia-Vidal, J.~C. Cuevas, E. Meyhofer, and P. Reddy, \emph{Radiative heat transfer in the extreme near field}, Nature \textbf{528}, 387 (2015).
\bibitem{StGelaisNatureNano16}R. St-Gelais, L. Zhu, S. Fan, and M. Lipson, \emph{Near-field radiative heat transfer between parallel structures in the deep subwavelength regime}, Nat. Nanotechnol. \textbf{11}, 515 (2016).
\bibitem{KloppstecharXiv}K. Kloppstech, N. K\"{o}nne, S.-A. Biehs, A. W. Rodriguez, L. Worbes, D. Hellmann, and A. Kittel, \emph{Giant heat transfer in the crossover regime between conduction and radiation}, Nat. Commun. \textbf{8}, 14475 (2017).
\bibitem{WatjenAPL16}J.~I. Watjen, B. Zhao, and Z.~M. Zhang, \emph{Near-field radiative heat transfer between doped-Si parallel plates separated by a spacing down to 200 nm}, Appl. Phys. Lett. \textbf{109}, 203112 (2016).
\bibitem{Thomas}N.~H. Thomas, M.~C. Sherrott, J. Broulliet, H.~A. Atwater, and A.~J. Minnich, \emph{Electronic Modulation of Near-Field Radiative Transfer in Graphene Field Effect Heterostructures}, Nano Letters \textbf{19}, 3898 (2019).
\bibitem{Papadakis}G.~T. Papadakis, B. Zhao, S. Buddhiraju, and S. Fan, \emph{Gate-Tunable Near-Field Heat Transfer}, ACS Photonics \textbf{6}, 709 (2019).
\bibitem{Chen}K. Chen, P. Santhanam, and S. Fan, \emph{Near-Field Enhanced Negative Luminescent Refrigeration}, Phys. Rev. Appl. \textbf{6}, 024014 (2016).
\bibitem{Zhu}L. Zhu, A. Fiorino, D. Thompson, R. Mittapally, E. Meyhofer, and P. Reddy, \emph{Near-field photonic cooling through control of the chemical potential of photons}, Nature \textbf{566}, 239 (2019).
\bibitem{Ekeroth}R.~M. Abraham Ekeroth, P. Ben-Abdallah, J.~C. Cuevas, and A. García-Martín, \emph{Anisotropic Thermal Magnetoresistance for an Active Control of Radiative Heat Transfer}, ACS Photonics \textbf{5}, 705 (2018).
\bibitem{Ilic}O. Ilic, N.~H. Thomas, T. Christensen, M.~C. Sherrott, M. Soljačić, A.~J. Minnich, O.~D. Miller, and H.~A. Atwater, \emph{Active Radiative Thermal Switching with Graphene Plasmon Resonators}, ACS Nano \textbf{12}, 2474 (2018).
\bibitem{Messina_prb2017}R. Messina, P. Ben-Abdallah, B. Guizal, and M. Antezza, \emph{Graphene-based amplification and tuning of near-field radiative heat transfer between dissimilar polar materials}, Phys. Rev. B. \textbf{96}, 045402 (2017).
\bibitem{Moncada}E. Moncada-Villa, V. Fernandez-Hurtado, F.~J. García-Vidal, A. García-Martin, and J.~C. Cuevas, \emph{Magnetic field control of near-field radiative heat transfer and the realization of highly tunable hyperbolic thermal emitters}, Phys. Rev. B \textbf{92}, 125418 (2015).
\bibitem{Cui}L.~J. Cui, Y. Huang, J. Wang, and K.Y. Zhu, \emph{Ultrafast modulation of near-field heat transfer with tunable metamaterials}, Appl. Phys. Lett. \textbf{102}, 053106 (2013).
\bibitem{Ge}L. Ge, Z. Xu, Y. Cang, and K. Gong, \emph{Modulation of near-field radiative heat transfer between graphene sheets by strain engineering}, Opt. Express \textbf{27}, A1109 (2019).
\bibitem{Ghanekar}A. Ghanekar, M. Ricci, Y. Tian, O. Gregory, and Y. Zheng, \emph{Strain-induced modulation of near-field radiative transfer}, Appl. Phys. Lett. \textbf{112}, 241104 (2018).
\bibitem{Kou}J. Kou and A. J. Minnich, \emph{Dynamic optical control of near-field radiative transfer}, Opt. Express \textbf{26}, A729 (2018).
\bibitem{Segal}D. Segal and A. Nitzan, \emph{Molecular heat pump}, Phys. Rev. E \textbf{73}, 026109 (2006).
\bibitem{Shuttling}I. Latella, R. Messina, J.~M. Rubi, and P. Ben-Abdallah, \emph{Radiative Heat Shuttling}, Phys. Rev. Lett. \textbf{121}, 023903 (2018).
\bibitem{Segal2}D. Segal, \emph{Stochastic Pumping of Heat: Approaching the Carnot Efficiency}, Phys. Rev. Lett. \textbf{101}, 260601 (2008).
\bibitem{NDTR}L. Zhu, C.~R. Otey, and S. Fan, \emph{Negative differential thermal conductance through vacuum}, Appl. Phys. Lett. \textbf{100}, 044104 (2012).
\bibitem{Kottos}H. Li, L.~J. Fernández-Alcázar, F. Ellis, B. Shapiro, and T. Kottos, \emph{Adiabatic Thermal Radiation Pumps for Thermal Photonics}, Phys. Rev. Lett. \textbf{123}, 165901 (2019).
\bibitem{Berry}M.~V. Berry and M. Wilkinson, \emph{Quantal Phase Factors Accompanying Adiabatic Changes}, Proc. R. Soc. A \textbf{392}, 15 (1984).
\bibitem{Mott}M.~M. Qazilbash, M. Brehm, B.~G. Chae, P.-C. Ho, G.~O. Andreev, B.~J. Kim, S.~J. Yun, A.~V. Balatsky, M.~B. Maple, F. Keilmann, H.~T. Kim, and D.~N. Basov, \emph{Mott Transition in VO$_2$ Revealed by Infrared Spectroscopy and Nano-Imaging}, Science \textbf{318}, 1750 (2007).
\bibitem{Baker}A.~S. Barker, H.~W. Verleur, and H.~J. Guggenheim, \emph{Infrared Optical Properties of Vanadium Dioxide Above and Below the Transition Temperature}, Phys. Rev. Lett. \textbf{17}, 1286 (1966).
\bibitem{vanZwol}P. van Zwol, K. Joulain, P. Ben-Abdallah, and J. Chevrier, \emph{Phonon polaritons enhance near-field thermal transfer across the phase transition of VO$_2$}, Phys. Rev. B \textbf{84}, 161413(R) (2011).
\bibitem{Heiss}W. Heiss, \emph{The physics of exceptional points}, J. Phys. A: Math. Theor. \textbf{45}, 444016 (2012).
\bibitem{PBA-PRL2011}P. Ben-Abdallah, S.-A. Biehs, and K. Joulain, \emph{Many-Body Radiative Heat Transfer Theory}, Phys. Rev. Lett. \textbf{107}, 114301 (2011).
\bibitem{KrugerPRB12}M. Kr\"{u}ger, G. Bimonte, T. Emig, and M. Kardar, \emph{Trace formulas for nonequilibrium Casimir interactions, heat radiation, and heat transfer for arbitrary objects}, Phys. Rev. B \textbf{86}, 115423 (2012).
\bibitem{MessinaNdipoles}R. Messina, M. Tschikin, S.-A. Biehs, and P. Ben-Abdallah, \emph{Fluctuation-electrodynamic theory and dynamics of heat transfer in systems of multiple dipoles}, Phys. Rev. B \textbf{88}, 104307 (2013).
\bibitem{Messina-PRA2014}R. Messina, M. Antezza, \emph{Three-body radiative heat transfer and Casimir-Lifshitz force out of thermal equilibrium for arbitrary bodies}, Phys. Rev. A \textbf{89}, 052104 (2014).
\bibitem{Latella}I. Latella, P. Ben-Abdallah, S.-A. Biehs, M. Antezza, R. Messina, \emph{Radiative heat transfer and nonequilibrium Casimir-Lifshitz force in many-body systems with planar geometry}, Phys. Rev. B \textbf{95}, 205404 (2017).
\bibitem{Pramod3body}D. Thompson, L. Zhu, E. Meyhofer, and P. Reddy, \emph{Nanoscale radiative thermal switching via multi-body effects}, Nat. Nanotechnol. \textbf{15}, 99 (2020).
\bibitem{Philchain}P. Ben-Abdallah, K. Joulain, J. Drevillon, and C.~L. Goff, \emph{Heat transport through plasmonic interactions in closely spaced metallic nanoparticle chains}, Phys. Rev. B \textbf{77}, 075417 (2008).
\bibitem{abajo}A. Manjavacas and F.~J. Garcia de Abajo, \emph{Radiative heat transfer between neighboring particles}, Phys. Rev. B \textbf{86}, 075466 (2012).
\bibitem{Thouless} D. J. Thouless, \emph{Quantization of particle transport}, Phys. Rev. B \textbf{27}, 6083 (1983).
\bibitem{Palik}\textit{Handbook of Optical Constants of Solids}, edited by E. Palik (Academic Press, New York, 1998).
\bibitem{Springer}I. Chasiotis, in \emph{Springer Handbook of Experimental Solid Mechanics}, edited by W.~N. Sharpe (Springer, New York, 2008).
\end{thebibliography}
\end{document}